\documentclass[11pt]{article} \usepackage{mymor,epsfig}

\def\Journal#1#2#3#4{{#1} {\bf #2}, #3 (#4)}

\def\PLB{{\em Phys. Lett.}  B}
\def\PRL{\em Phys. Rev. Lett.}

\def\PRC{{\em Phys. Rev.} C}

\def\HIP{\em Heavy Ion Physics}

\def\be{\begin{equation}}
\def\ee{\end{equation}}
\def\bea{\begin{eqnarray}}
\def\eea{\end{eqnarray}}

\def\Rs{R_{\rm side}}

\def\la{\left \langle}
\def\ra{\right \rangle}
\begin{document}
\vspace*{4cm}
\title{BLAST-WAVE SNAPSHOTS FROM RHIC}

\author{BORIS TOM\'A\v SIK}

\address{CERN, TH Division, CH-1211 Geneva 23, Switzerland}

\maketitle\abstracts{%
I present fits with the so-called blast-wave model to single-particle 
spectra and HBT correlations from Au+Au collisions at a CMS energy of 
130 $A$GeV. There is only  little choice of freeze-out temperature 
and transverse flow velocity for which the model fits both the identified
spectra and the correlation radii just well enough
not to be excluded. The observed steep $M_\perp$ dependence of 
$R_{\rm side}$ leads to a temperature which it is problematic to interpret.
The applicability of the model for the freeze-out description is thus
questioned.
}


\section{HBT interferometry in heavy-ion collisions}

In heavy-ion collisions we study the collective behaviour of 
strongly interacting matter. HBT interferometry is a method that 
helps us to determine the final state of the fireball evolution,
the so-called {\em freeze-out}. We thus obtain a snapshot of the 
result to which the collective evolution of the fireball leads.

A particularly interesting phenomenon at the freeze-out is 
the {\em transverse expansion}, as this is not a part of the 
initial conditions and is entirely generated by pressure
of the QCD matter. Another interesting quantity is the {\em freeze-out
temperature}, which characterizes 
the end of the collective system evolution.
It has been argued \cite{CL} that both these quantities can be 
determined unambiguously from single-particle $p_\perp$ spectra 
and two-particle HBT correlations. 

Here I report on such a project 
in the framework of the so-called {\em blast-wave model}. I analysed
identified single-particle spectra \cite{phspec} and HBT correlation
radii \cite{phhbt,sthbt} from central Au+Au collisions at a RHIC energy
of $\sqrt{s} = 130\, A\mbox{GeV}$.


\section{The (blast-wave) model}

The main assumptions of this---now widely used---model, which are relevant to 
this study are:
\begin{enumerate}
\item 
Pions, nucleons and also kaons decouple all quite {\em suddenly} 
from the whole transverse profile of the fireball. 
For all of them the freeze-out
happens at {\em the same proper time}, measured in a frame that 
co-moves longitudinally with the fluid element of the expanding fireball.
\item 
The radial density distribution at the freeze-out is {\em uniform}.
\item 
Longitudinal expansion is {\em boost-invariant}. Heavy-ion experts
know this as Bjorken scenario, the rest of the world is familiar with
the astrophysical analogue: the Hubble expansion.
\item 
In this study, the {\em transverse expansion} is parametrized through
rapidity, which depends linearly on the radial coordinate.
\end{enumerate}
Technically, these assumptions are expressed through the emission 
function \cite{fitpap}, which is the Wigner density of 
the source normalized to the number of particles
\bea
\nonumber
S(x,p) \, d^4x & = & 
\frac{1}{(2\pi)^3}\,  m_\perp\cosh(y-\eta)\, 
\exp \left (-\frac{p_\mu u^\mu - \mu}{T}\right )\,
\theta(R_B - r) \label{ef}\\
&&
\frac{1}{\sqrt{2\pi\Delta\tau^2}}\, 
\exp\left ( - \frac{(\tau - \tau_0)^2}{2\Delta\tau^2}\right ) \,
\tau \, d\tau \, d\eta\, r\, dr\, d\phi\, , \\
u^\mu & = & (\cosh\eta_t\, \cosh\eta,\, \sinh\eta_t\, \cos\phi,\,
\sinh\eta_t\, \sin\phi,\, \cosh\eta_t\, \sinh\eta)\, ,
\\
\eta_t & = & \sqrt{2}\, \eta_f\, \frac{r}{R_B}\, .
\eea
In this notation, space-time coordinates and the momentum in the 
so-called {\em out-side-long} system are parametrized as
\bea
x^\mu & = & 
(\tau\, \cosh\eta,\, r\, \cos\phi,\, r \sin\phi,\, \tau\, \sinh\eta)\, \\
p^\mu & = & (m_\perp \, \cosh y,\, p_\perp,\, 0,\, m_\perp\, \sinh y)\, .
\eea
Model parameters are to be determined from a fit to data. These include:
temperature $T$, scaled transverse flow gradient $\eta_f$, transverse 
geometric radius $R_B$, mean Bjorken lifetime  $\tau_0$,
and mean proper emission duration $\Delta\tau$. 
The chemical potential $\mu$ is not being determined in this study.
For the presentation of
the results the average transverse velocity is used 
\be
\la v_t \ra = \frac{2}{R_B^2} \, {\int_0^{R_B} r\,  dr \, \tanh \eta_t(r)}\, .
\ee

For simplicity, Boltzmann distribution has been used in
Eq.~\ref{ef}. Note that the model is formulated as thermal: it is assumed
that particles decouple from a system in local thermal equilibrium with 
the temperature $T$. In the corresponding term, the particle momentum is 
coupled to the local flow velocity as $p_\mu u^\mu$ in order to 
obtain the energy in the rest frame of the fluid. The strength of this coupling
is controlled by the temperature. The lower the temperature, the stronger
the momentum of the particle corresponds to the fireball expansion 
velocity and the more pronounced the effects of the expansion 
in the observables are. In terms of HBT radii, the expansion is encoded in
their $M_\perp$ dependence. A lower value of the temperature parameter 
thus leads to a stronger $M_\perp$ dependence.

Single-particle spectra were calculated via \cite{CL}
\be
E_p \frac{dN}{d^3p} =  \int d^4x \, S(x,p) \, .
\ee
The HBT correlation radii were obtained from a numerical evaluation of 
the model-independent expressions \cite{CSH}, in which the 
second spatial moments of the emission function are used.


\section{Fits to (low-momentum) single-particle $p_\perp$ spectra}

With the blast-wave model, I fitted single-particle spectra of identified 
positive and negative pions, kaons and protons as measured by the 
PHENIX Collaboration \cite{phspec}. Bose--Einstein statistics and resonance 
decays were assumed for pions. I assumed baryon chemical potential for the 
resonances as in an earlier paper \cite{fitpap}, but no pion chemical 
potential was included. 

An important issue in the analysis is that every spectrum was fitted 
individually. This allows for a check of the assumption that all particles
freeze-out simultaneously. If so, fits to different spectra 
would lead to compatible results \footnote{The slope of the spectrum is 
determined by the temperature, the strength of the transverse expansion, 
and the mass of the particles. \cite{SSH}}. On the other hand, if the results 
do not agree, the assumption is wrong. 

\begin{figure}
\begin{minipage}{8cm}
\begin{center}
\setlength{\epsfxsize}{7.7cm}
\epsffile{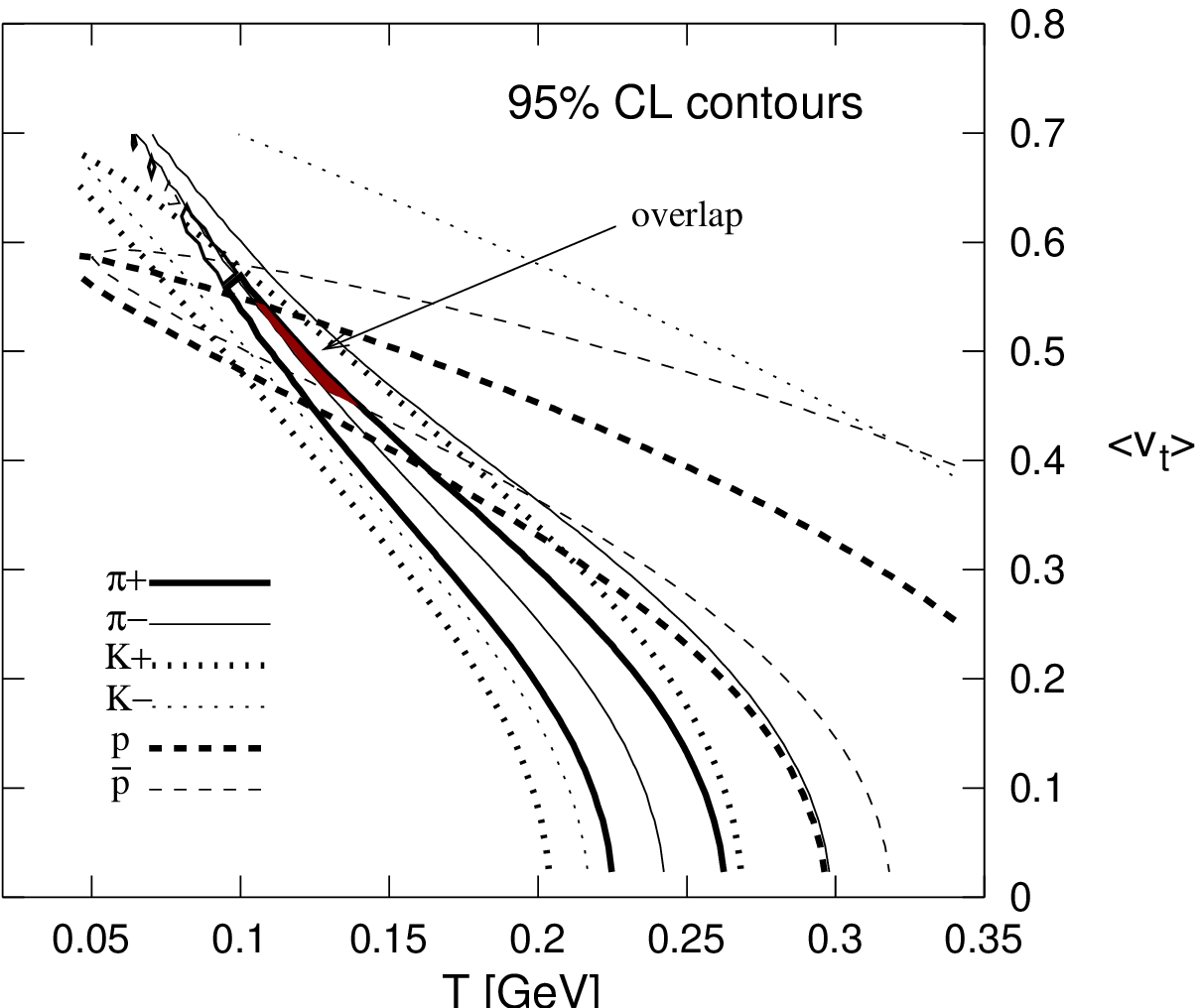}
\end{center}
\end{minipage}
\begin{minipage}{8cm}
\begin{center}
\setlength{\epsfxsize}{7.7cm}
\setlength{\epsfysize}{6.9cm}
\epsffile{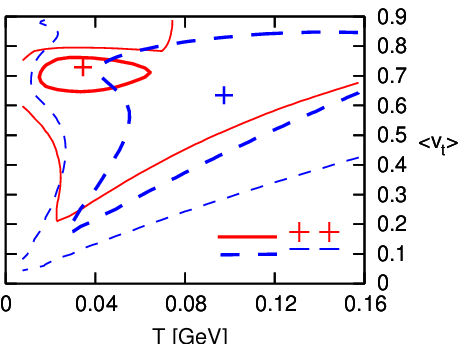}
\end{center}
\end{minipage}
\caption{%
Left: The 95\% confidence level contours in temperature and average transverse 
flow velocity resulting from fits to identified single-particle spectra. 
Right: The 1$\sigma$ (thick lines) and 95\% confidence level (thin lines) 
contours from fits to HBT radii from both STAR \protect\cite{sthbt} and PHENIX
\protect\cite{phhbt}. Solid lines are for $\pi^+\pi^+$ correlations, 
dashed lines for $\pi^-\pi^-$ correlations. Crosses denote the position
of the best fits.}
\label{f:contours}
\end{figure}
%
There is no overlap between the fit results to different spectra 
at the 1$\sigma$ level. Can the model be ruled out? In order to find out,
I plot the contours corresponding to 95\% confidence levels 
from the fits in Fig.\ \ref{f:contours}. An overlap is found at this 
level, hence the model is not ruled out by the fits to spectra. 

It remains to be checked whether the quality of the fits can be improved
by fine-tuning the details of the model: changing the radial dependence
of the transverse rapidity, introducing pion chemical potential, etc.


\section{Fits to HBT correlation radii}

\begin{figure}
\vspace*{-1cm}
\setlength{\epsfysize}{10cm}
\centerline{{\epsffile{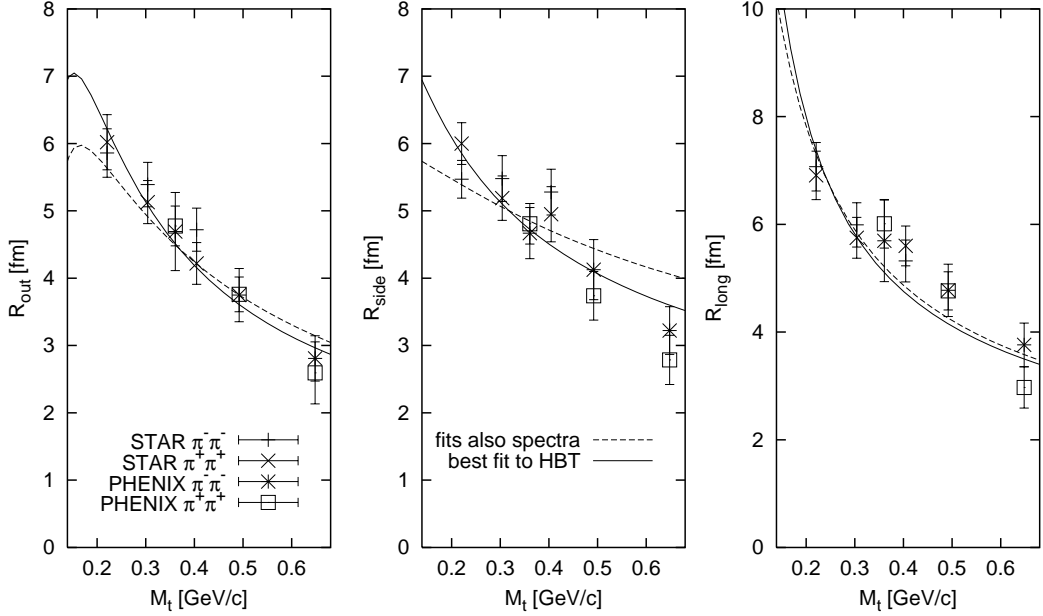}}}
\vspace*{-0.6cm}
\caption{%
Example fits to HBT radii measured at RHIC. Dashed lines show the best model 
which fits both spectra and HBT: $T=106\, \mbox{MeV}$,
$\la v_t\ra = 0.53$, $R_B = 12.36\, \mbox{fm}$, $\tau_0 = 4.54\, \mbox{fm}/c$,
$\Delta\tau = 4.57\,\mbox{fm}/c$. Solid lines correspond to the best fit
to $\pi^+\pi^+$ correlations: $T=33\, \mbox{MeV}$, $\la v_t \ra = 0.73$,
$R_B = 24.11\, \mbox{fm}$, $\tau_0 = 21.32\, \mbox{fm}/c$,
$\Delta\tau = 1.09\, \mbox{fm}/c$.
}
\label{f:exam}
\end{figure}
%
The measurements of HBT radii by STAR and PHENIX cover different $M_\perp$
regions, with only one data point overlapping 
(Fig.\ \ref{f:exam}). The PHENIX data
show a steeper $M_\perp$ dependence of $\Rs$ than those of STAR.
Such a steep $\Rs(M_\perp)$ would ask for a strong
transverse flow and a low temperature. 

Indeed, this is confirmed by the fits. At the 1$\sigma$ level, 
some results from 
fitting data sets from the two collaborations do not agree. In order to 
have a robust statement about whether the model fails to reproduce the
data, systematical errors  quoted by the experiments were {\em added 
linearly} to the statistical ones. Under these circumstances one finds 
a large  overlap at 95\% confidence level from fitting all four data sets.

In order to improve statistics, data of 
the same charge from both collaborations
were added together and fitted. Resulting $\chi^2$ contour plots are
displayed in Fig.\ \ref{f:contours}. Note that there is only a tiny overlap
between the 95\% CL contour of $\pi^+\pi^+$ correlations with the result
of fitting single-particle spectra. It is located at 
$T \approx 106\, \mbox{MeV}$.
Furthermore, the best fit to $\pi^+\pi^+$ correlations is obtained at 
$T=33\, \mbox{MeV}$ and $\la v_t\ra = 0.73$! This is not to be interpreted 
at the real physical freeze-out temperature! As seen from Fig.\ \ref{f:exam},
these values of the model parameters are required in order to produce the
observed steep $M_\perp$ dependence of $\Rs$. Thus $T$ is merely to 
be interpreted as a parameter that controls the coupling of momentum to 
expansion velocity in the framework of the blast-wave model.

It is interesting to note that similar results
appear from fitting the HBT data from the SPS program and the preliminary
data from the RHIC run at full energy. This study is in progress.


\section{Conclusions}

Summarizing the main observations: first, the blast-wave model can
fit the spectra {\em and} the correlation radii only marginally. The resulting
parameters are close to the 95\% CL contour of both fits. Second, the best
fit to HBT radii is achieved with model parameters that are hard to 
interpret phenomenologically. My conclusion is that the blast-wave model
is probably {\em not a suitable description of the freeze-out}. 

Note that we do have indications from cascade generators \cite{bleicher} 
and studies of pion scattering rate \cite{tw}, which show that the freeze-out
takes place continuously and there may even be ordering in the production of 
different species and transverse momenta. This feature is in contrast to 
the simple assumption of the blast-wave model, which says that all particles
freeze out suddenly at the same time. 

It will be crucial to formulate a good description of the freeze-out.
This is because the momentum spectra are produced at freeze-out. If these
spectra are to be searched for signatures of collective behaviour, 
it is important to understand the process in which they are produced.


\end{document}